\UseRawInputEncoding 
\documentclass[%
 reprint,
superscriptaddress,
 amsmath,amssymb,
 aps,
prm,
floatfix,
]{revtex4-2}
\usepackage{hyperref}
\usepackage{graphicx}
\usepackage{dcolumn}
\usepackage{bm}
\usepackage{color}
\usepackage{float}
\usepackage{xfrac}

\begin{document}

\title{Structural changes induced by electric currents in a single crystal of Pr$_2$CuO$_4$}

\author{Susmita Roy}
\affiliation{Department of Physics, University of Colorado at Boulder, Boulder, Colorado 80309, USA}

\author{Feng Ye}
\affiliation{Neutron Scattering Division, Oak Ridge National Laboratory, Oak Ridge, TN, 37830, USA}

\author{Zachary Morgan}
\affiliation{Neutron Scattering Division, Oak Ridge National Laboratory, Oak Ridge, TN, 37830, USA}

\author{Kabir Mathur}
\affiliation{Department of Physics, University of Colorado at Boulder, Boulder, Colorado 80309, USA}

\author{Anish Parulekar}
\affiliation{Department of Physics, University of Colorado at Boulder, Boulder, Colorado 80309, USA}

\author{Syed I. A. Jalali}
\affiliation{Materials Science and Engineering Program, Department of Mechanical Engineering, University of Colorado Boulder, Boulder,CO 80309,USA.}

\author{Yu Zhang}
\affiliation{Department of Physics, University of Colorado at Boulder, Boulder, Colorado 80309, USA}

\author{Gang Cao}
\affiliation{Department of Physics, University of Colorado at Boulder, Boulder, Colorado 80309, USA}

\author{Nobu-Hisa Kaneko}
\affiliation{National Institute of Advanced Industrial Science and Technology (AIST), Tsukuba, Ibaraki 305-8563, Japan}

\author{Martin Greven}
\affiliation{School of Physics and Astronomy, University of Minnesota, Minneapolis, Minnesota 55455, USA}

\author{Rishi Raj}
\affiliation{Materials Science and Engineering Program, Department of Mechanical Engineering, University of Colorado Boulder, Boulder,CO 80309,USA.}

\author{Dmitry Reznik}
\affiliation{Department of Physics, University of Colorado at Boulder, Boulder, Colorado 80309, USA}
\affiliation{Center for Experiments on Quantum Materials, University of Colorado at Boulder, Boulder, Colorado 80309, USA}

\date{\today}

\begin{abstract}
We demonstrate a novel approach to the structural and electronic property modification of perovskites, focusing on Pr$_2$CuO$_4$, an undoped parent compound of a class of electron-doped copper-oxide superconductors. Currents were passed parallel or perpendicular to the copper-oxygen layers with the voltage ramped up until a rapid drop in the resistivity was achieved, a process referred to as "flash". The current was then further increased tenfold in current-control mode. This state was quenched by immersion into liquid nitrogen. Flash can drive many compounds into different atomic structures with new properties, whereas the quench freezes them into a long-lived state. Single-crystal neutron diffraction of as-grown and modified Pr$_2$CuO$_4$ revealed a $\sqrt{10}$x$\sqrt{10}$ superlattice due to oxygen-vacancy order. The diffraction peak intensities of the superlattice of the modified sample were significantly enhanced relative to the pristine sample. Raman-active phonons in the modified sample were considerably sharper. Measurements of electrical resistivity, magnetization and two-magnon Raman scattering indicate that the modification affected only the Pr-O layers, but not the Cu-O planes. These results point to enhanced oxygen-vacancy order in the modified samples well beyond what can be achieved without passing electrical current. Our work opens a new avenue toward electric field/quench control of structure and properties of layered perovskite oxides.
\end{abstract}

\maketitle

\section{Introduction}
It was recently demonstrated that the application of a moderate electric field (E-field) at elevated temperatures can modify structural, electrical, optical, and other properties of a vast majority of crystalline and polycrystalline materials, effectively turning them into new materials.\cite{cologna2010flash} For example, when a $\sim$100 V/cm E-field is applied to a single crystal of rutile, TiO$_2$, and the  material is very briefly heated far above room temperature, its electrical conductivity increases dramatically and the material begins to glow in an electroluminescence (EL) like phenomenon referred to as ``flash" \cite{yoon2018measurement}. Materials revert to their original state when the E-field is removed, but the new state may be effectively frozen-in via a liquid-nitrogen quench. This phenomenon in polycrystalline materials was originally attributed to Joule heating at grain boundaries \cite{yoon2018measurement}, but its discovery in single crystals casts doubt on this model. This is a very active area of basic science as well as technology (e.g., ceramics sintering \cite{todd2015electrical,yu2017review}). There exist two main challenges: (i) to understand the underlying microscopic mechanisms, and (ii) to learn how to harness the potential of this approach to create new materials. The results presented here address the second challenge.

Compelling evidence exists that materials in the flash state adopt new crystal structures as well as local structures characterized by interstitial defects. For example, in-situ experiments x-ray diffraction experiments demonstrated the emergence of a new phase (identified to be pseudo-cubic) during the flash state evidenced by an appearance and disappearance of a new Bragg peak \cite{Lebrun2015}.

Whereas the existence of such new states has been demonstrated, the extent to which these structures can be quenched down to ambient conditions is not completely certain. To the best of our knowledge all previous work in this area was done on band insulators or metals. Here we propose to apply the flash-quench technique to materials whose physical properties are very sensitive small perturbations. 

Mott or charge-transfer insulators are prototypical examples of such materials because they are undergo phase transitions that can be controlled by doping, external fields, disorder and temperature \cite{Keimer2015}. Our overarching goal is to establish how the local atomic structure of these phases can be manipulated through the flash/quench combination, and what effect this has on magnetic, electric, thermal, and spectroscopic properties. As a first step in this direction, we examined the effects of flash on a single crystal of the charge-transfer insulator Pr$_2$CuO$_4$ (PCO), which is well known as a parent compound, which becomes an electron-doped electronically-correlated high temperature superconductor upon chemical substitution or changes in the oxygen stoichiometry.

\begin{figure}
\includegraphics[width=0.5\textwidth]{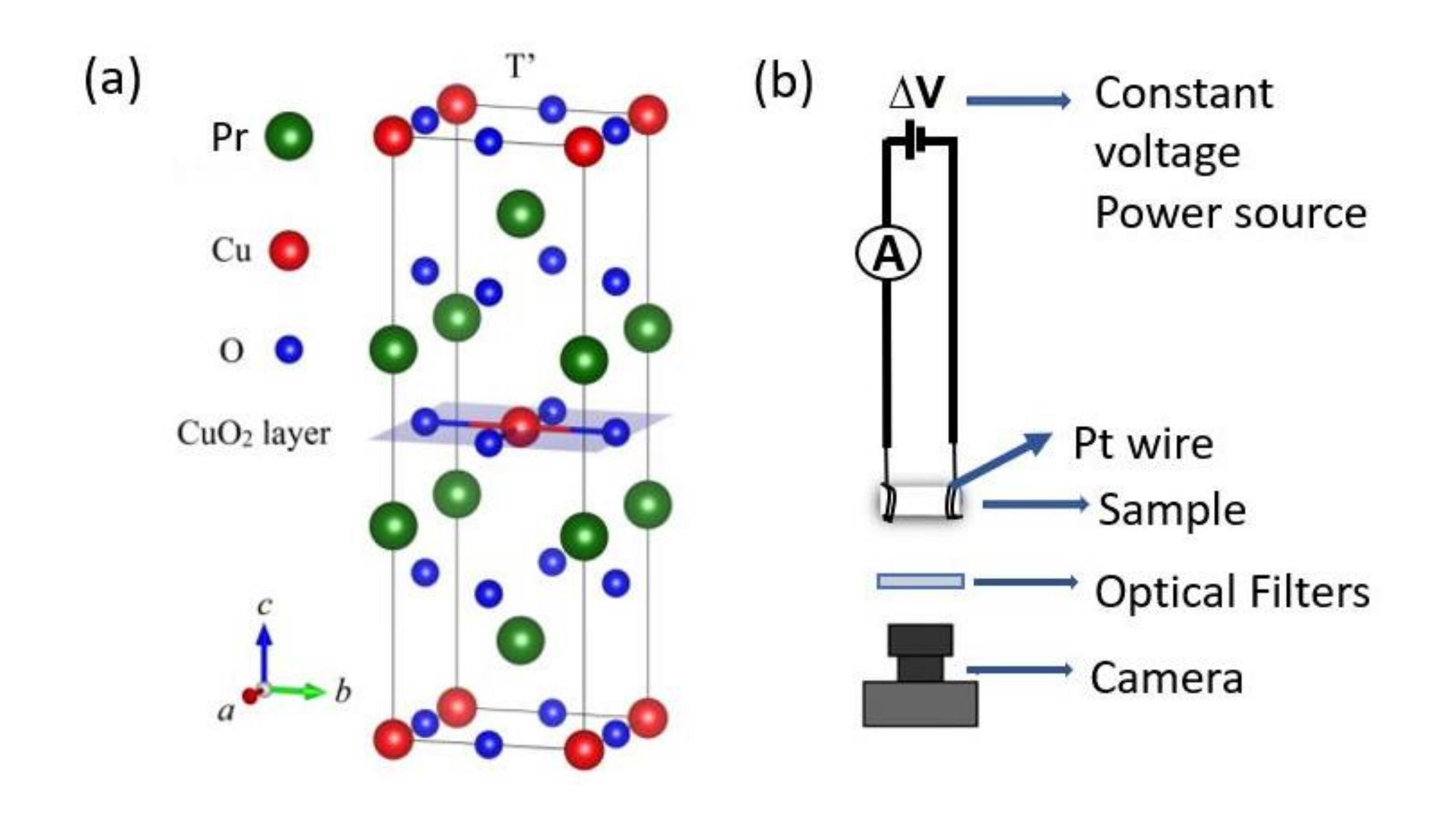}
\centering
    \caption{(a) The T' structure of PCO. (b) Schematic of the flash experiment. }
    \label{fig:Flash_schematic}
\end{figure}

The effect of flash has been examined on powders, polycrystals, and single crystals of a variety of oxides \cite{cologna2010flash, cologna2011flash,yoshida2014densification, karakuscu2012defect, biesuz2016flash, muccillo2014electric,yadav2017onset,kathiria2022flash}, but to the best of our knowledge, not yet on electronically-correlated layered perovskites. PCO crystallizes in the T' structure, consisting of alternating CuO$_2$ and PrO planes (Fig. \ref{fig:Flash_schematic}a) \cite{sanjuan1995raman,armitage2010progress,allenspach1989magnetic}. The CuO$_2$ plane is a square lattice in which each Cu atom is surrounded by four neighboring O atoms, and apical O atoms above or below the plane are nominally absent. This structure is contrasted by the tetrahedral or octahedral coordination of Cu and O atoms in the T structure of hole-doped superconductors and their parent compounds \cite{tokura1989superconducting}. Compared to the T structure, the T' structure of PCO exhibits a shorter interlayer distance and longer Cu-Cu distance \cite{muller1975ternare}. The CuO$_2$ plane is defined as the \textit{ab}-plane, and the \textit{c}-axis is perpendicular to this plane. PCO is a charge-transfer insulator \cite{homes2002infrared} in which the planar Cu-O superexchange interaction induces antiferromagnetism. The interplane exchange coupling along the $c$-direction is much weaker and frustrated.

Our original interest was to see if flash followed by a liquid nitrogen (LN$_2$) quench could be used to change the electronic structure of the copper oxygen planes. We therefore focused on probing electrical transport, magnetic susceptibility, as well as two-magnon Raman scattering as a probe known to be sensitive to carrier doping. In addition, examining the Raman scattering phonon spectrum can provide important clues about possible modifications of local atomic structure. We also used single-crystal x-ray and neutron diffraction to understand the effect of flash on the atomic structure. We uncovered that oxygen vacancies in PCO order into a superlattice already in the unmodified state, and that this order is enhanced by flash, with the effect of altering the physical properties of the PrO layers, while keeping the vibrational dynamics, electrical conductivity, and magnetic properties of the CuO$_2$ layers relatively unperturbed.

\section{Experimental details}

Flash consists of three stages. Typically the sample is initially kept in a furnace above room temperature. Stage I is the incubation, when the conductivity of the sample increases very slowly as the electric field is increased.  Stage II begins when the conductivity starts to increase abruptly in a non-linear fashion at a particular electric field strength, signaling the onset of flash. After this onset, the power supply is switched from voltage control to current control. Usually, a sample starts to glow brightly due to electroluminescence. It is possible that some samples do not visibly glow, which depends on the luminescence spectrum and brightness. Stage III is a current-controlled steady state where the sample stays in the active flash state. During this stage, the sample can be pulled out of the furnace without interrupting flash. Our experiment was done in air at room temperature, and then the sample was dipped into LN$_2$. Turning off the current while immersed in LN$_2$, quenches (i.e. freezes in) the atomic structure created during stage III. We found that PCO flashes at room temperature, so our experiment was carried out outside of a furnace at room temperature.

The single crystal was grown in a floating zone image furnace \cite{PhysRevB.70.094507}. We cut it into four orthorhombic pieces. Two pieces had the longest dimension parallel to the \textit{c}-axis (C-crystals), and the other two had the longest dimension parallel to the \textit{a}-axis (A-crystals). Among these four pieces, one A-crystal and one C-crystal were chosen for flash. We call these "flashed" samples. The other two pieces were not flashed, and are refered to as "unflashed" throughout the text. For the flash experiment, thin Pt wires serving as electrical connections were wrapped around both ends of the rods, and Pt paste was applied over the contacts for improved electrical connection (shown in Fig. \ref{fig:Flash_schematic}). The flashed C-crystal (Sample 1) has a length of 1.7 mm  and a cross section of 1.07 mm x 1.09 mm. The length of the flashed A-crystal (Sample 2) was 2.15 mm and the cross section of 1.7 mm x 0.81 mm. The onset of flash for Sample 1 was at 150 V/cm at current density of 100 mA/mm$^2$. After the onset, the current density kept increasing up to 1000 mA/mm$^2$ where this sample started to glow brightly. We held the crystal for 5 minutes under these conditions and then immersed it into LN$_2$. Sample 2 flashed at an electric field of 40 V/cm and current density of 100 mA/mm$^2$. Then the current increased to 1040 mA/mm$^2$ and the sample was immersed into LN$_2$. Sample 2 did not glow even at this higher current density. We kept both samples inside LN$_2$ for several minutes before performing other measurements. 

Standard four-probe electrical resistivity measurements were carried out on both unflashed and flashed samples in a closed-cycle cryostat (Advanced Research Systems). \textit{c}-axis resistivity was measured on the C-crystals, whereas the \textit{a}-axis resistivity was measured on the A-crystals. The C-crystal shattered after these measurements, during heating to room temperature, so only the A-crystals were used for subsequent measurements. Magnetic properties were measured using a Quantum Design (QD) MPMS-7 SQUID magnetometer.  The \textit{ab}-face of the unflashed and flashed A-crystals were used for  Raman scattering measurements to observe phonons and two-magnon scattering. The temperature dependence of the Raman spectra was measured by mounting the samples in a top loaded closed-cycle refrigerator, using a 532 nm laser and a single-stage McPherson spectrometer, equipped with a liquid nitrogen cooled charge-coupled device (CCD) detector and an 1800 grooves/mm grating. The entrance slit of the spectrometer was opened to be larger than the image of the laser spot to avoid chromatic aberrations of the collecting optics. Raman spectra were obtained for the four polarization geometries XX, XY, X'Y', X'X' where the XX/XY notation denotes that the incident laser polarization is parallel to the crystal axes (\textit{a} and \textit{b}), and the scattered light polarization is parallel/perpendicular to the incident laser polarization respectively. The X' and Y' directions are rotated 45$^{\circ}$ in the \textit{ab}-plane with respect to \textit{a} and \textit{b}. Two-magnon measurements at room temperature were performed in a Spex triple-spectrometer equipped with a water cooled CCD detector. Laser lines 476.5 nm and 496.5 nm from an Ar ion laser were used to clearly see the two-magnon peak. For the two stages 300 grooves/mm gratings were used whereas the final stage had a 150 grooves/mm grating. The data were corrected for the spectral response of the equipment using a calibrated lamp with a broad spectrum. We collected the Raman data from several different spots on each crystal and found no observable variation from spot to spot. The resistivity and magnetization were measured repeatedly on the same crystals and gave the same results every time.

The crystalline quality and stoichiometry of the A-crystals were characterized using x-ray single-crystal diffraction at the Rigaku XtaLAB PRO diffractometer housed at the Spallation Neutron Source (SNS) at Oak Ridge National Laboratory. The crystals were suspended in Paratone oil and mounted on a plastic loop attached to a copper pin/goniometer. The single-crystal x-ray diffraction data were collected with molybdenum K-$\alpha$ radiation ($\lambda = 0.71073~\AA$) at 293 K. Around 4000 diffraction Bragg peaks were indexed for both flashed and unflashed single crystals. Further characterization of the flashed and 
unflashed A-crystal samples at 200~K was carried out at the CORELLI diffracometer located at the Spallation Neutron Source \cite{ye18}. A 360$^{\circ}$ rotation scan was performed for both samples to map out the reciprocal space volumes using 360 sample orientations at 1~min/angle rate.

\section{Results}

\begin{figure}
\includegraphics[width=0.5\textwidth]{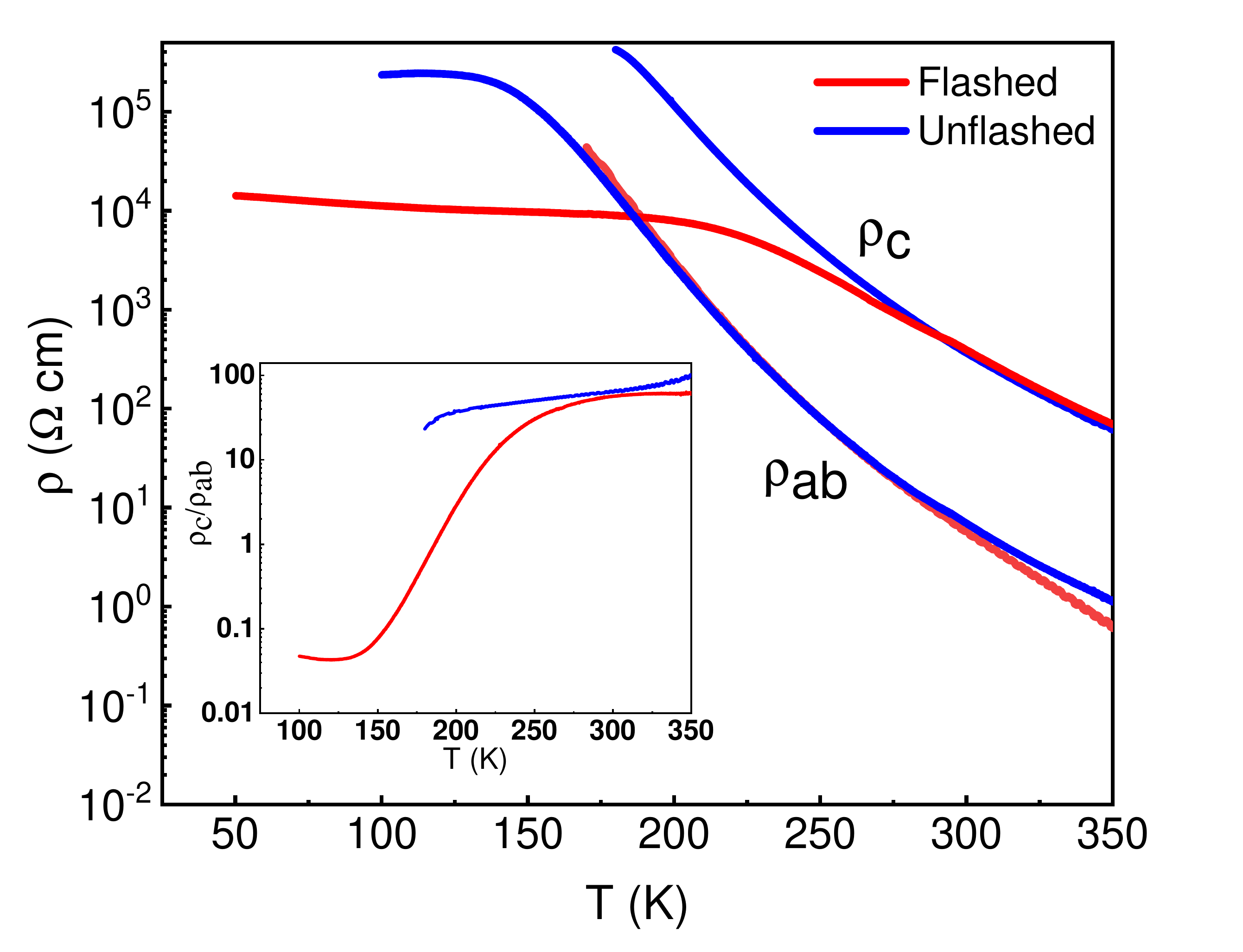}
\centering
    \caption{Temperature dependence of the electrical resistivity measured along the \textit{ab}-plane ($\rho_{ab}$) and along c direction ($\rho_{c}$) for both unflashed and flashed Pr$_2$CuO$_4$ samples. Inset shows the ratio of $\rho_{c}$ to $\rho_{ab}$.}
    \label{fig:ResistivityvsT}
\end{figure}

Figure \ref{fig:ResistivityvsT} shows how the flash experiment affects the \textit{ab}-plane and \textit{c}-axis resistivities of PCO in the temperature range 50 K to 350 K. The unflashed sample is an insulator, which is apparent from our data. $\rho_{c}$ of the flashed sample starts to decrease below 250 K and reaches nearly 10$^4$ $\Omega$ cm at 50 K. We were unable to measure the resistivity of the unflashed sample below 180 K because it is too high. The $a$-axis resistivity data for the unflashed sample are consistent with previous work \cite{homes2002infrared}. \textit{c}-axis resistivity data for PCO have not been previously reported, to the best of our knowledge. The inset in Fig. \ref{fig:ResistivityvsT} highlights the drastic change in the anisotropy of the resistivity between the flashed and unflashed samples, with the ratio between the \textit{c} and \textit{ab} resistivities decreasing sharply below 250K.

The key result is that flash reduces the low-temperature resistivity in the \textit{c}-direction by 2-3 orders of magnitude at lower temperatures. On the other hand, $\rho_{ab}$ of the flashed sample is almost the same  as for the unflashed one. 

\begin{figure}
\includegraphics[width=0.55\textwidth]{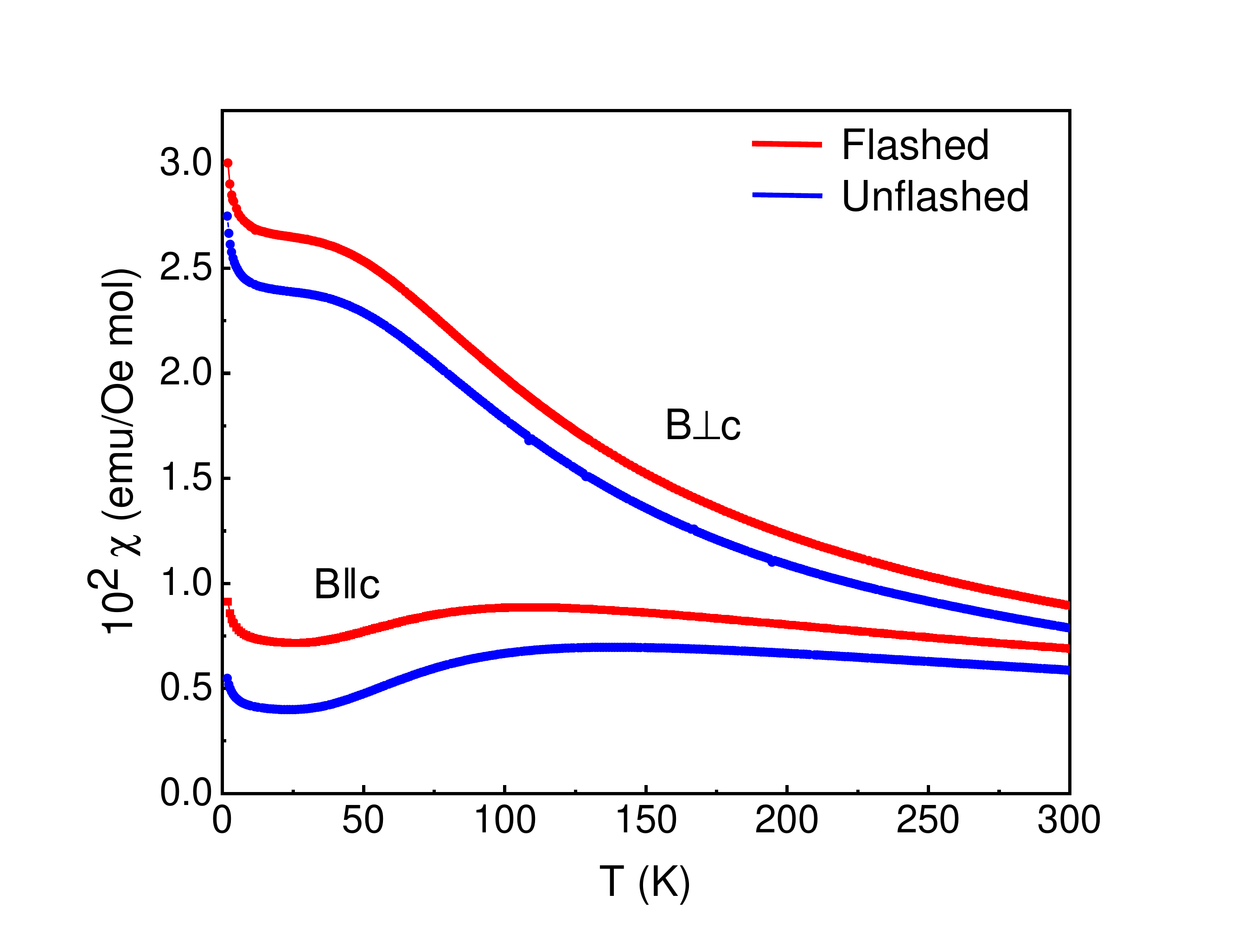}
\centering
    \caption{Temperature dependence of magnetic susceptibility of the unflashed and flashed Pr$_2$CuO$_4$ samples 
    measured with magnetic field parallel and perpendicular to the \textit{c}-direction.}
    \label{fig:MT}
\end{figure}

Figure \ref{fig:MT} shows the temperature dependence of the magnetic susceptibility of unflashed and flashed samples obtained with a small applied magnetic field of 100 Oe along and perpendicular to the \textit{c}-direction. The susceptibility of the flashed sample follows the behavior of the unflashed sample, but with an increased strength, reaching close to a factor of two at low temperatures in both directions. The unflashed result is close to previously published data. \cite{allenspach1989magnetic,foldeaki1991magnetic,hundley1989specific}.

\begin{figure}
\includegraphics[width=0.5\textwidth]{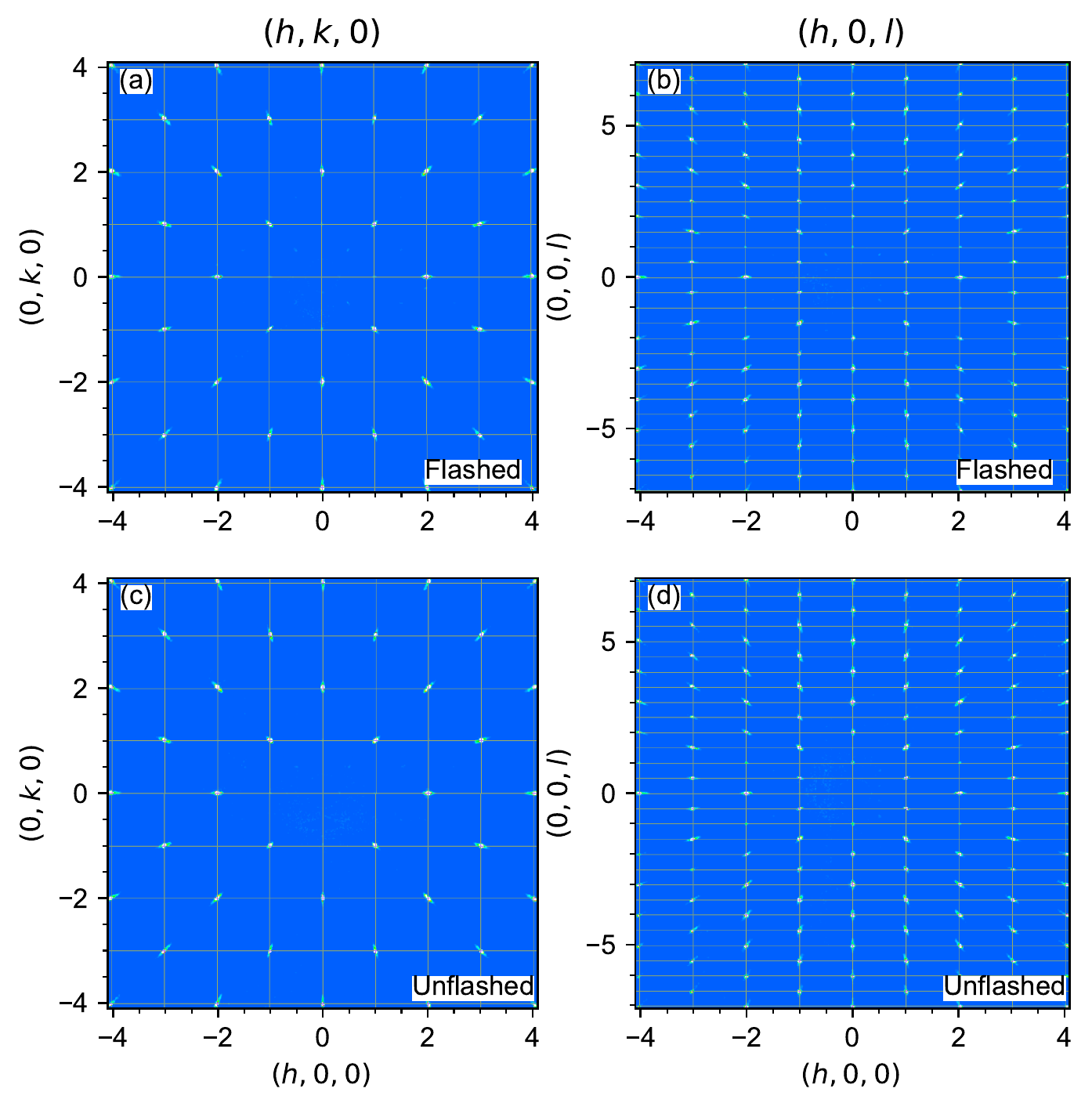}
\centering
    \caption{Representative single-crystal x-ray diffraction data of the flashed (a) $hk0$, 
    (b) $h0l$ and reference (un-flashed) (c) $hk0$, (d) $h0l$ samples. There is no indication of impurity or superlattice reflections.}
    \label{fig:Xray}
\end{figure}

The crystal structures obtained from x-ray diffraction show that both flashed and unflashed samples have the reported underlying body-centered structure ($I4/mmm$ space group), with lattice constants 
$a=b=3.96~\AA$ and $c=12.21~\AA$.  Figure \ref{fig:Xray} compares data for both samples in the $(hk0)$ and $(h0l)$ scattering planes. The reflections are refined using least-squares analysis, which 
indicates that the Pr and Cu sites are fully occupied and that the occupancy at the oxygen site of both samples is 95\%.

\begin{figure}
\includegraphics[width=0.5\textwidth]{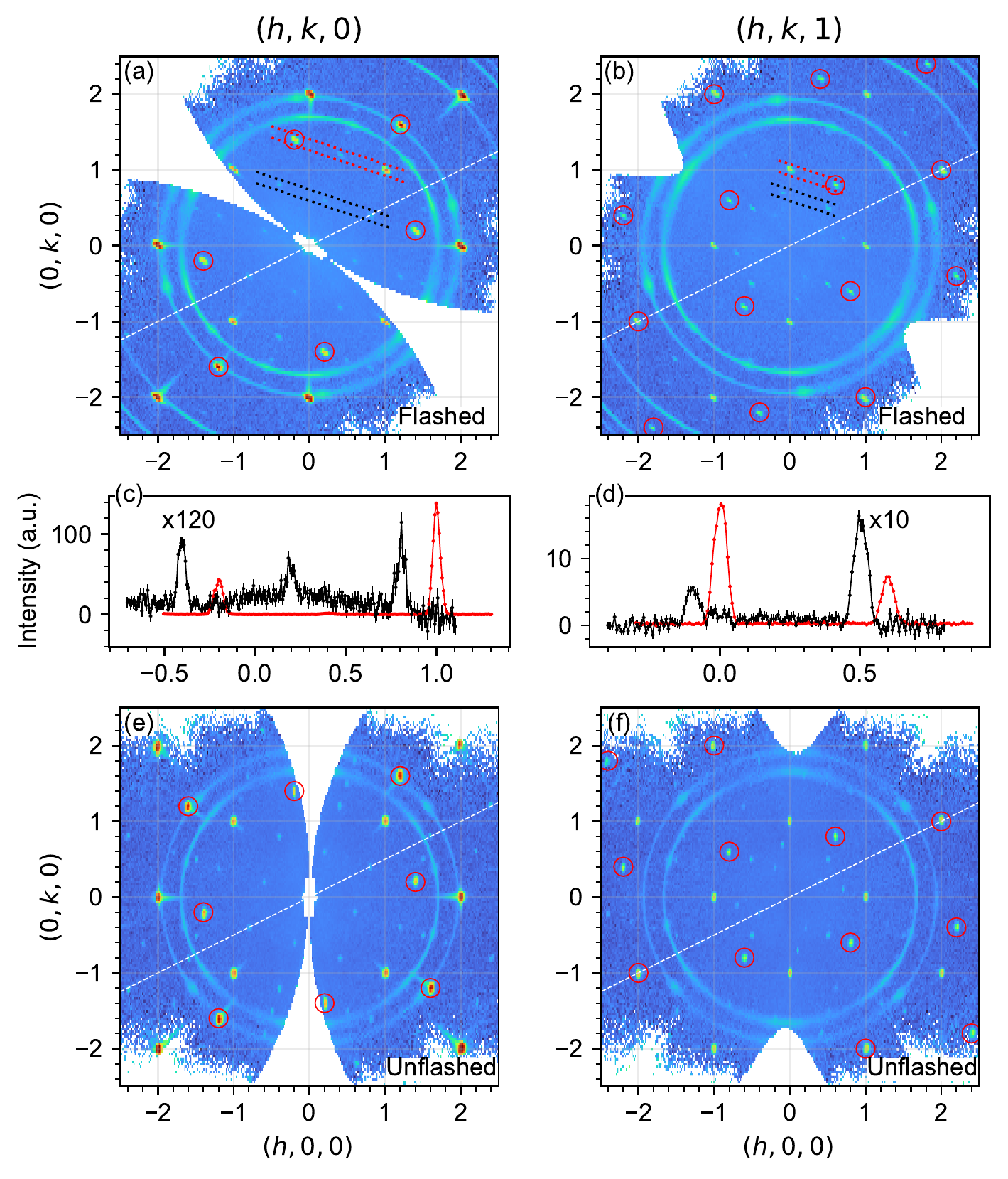}
\centering
    \caption{Map of single-crystal neutron diffraction data of a flashed single crystal in the $(hk0)$ (a) and $(hk1)$ (b) plane.  The dashed line indicates the 2-fold rotation axis where a twinned crystal is present. The red circles represent the nuclear Bragg peaks from the twinned domain. The volume fraction ratio between the twins is 2:1. (c,d) Line cuts across the nuclear and superlattice peaks along the red and black dashed lines in (a,b) projected on the $[1,0,0]$ direction. The superlattice intensity is scaled accordingly. (e,f) Data for the unflashed samples in the same scattering planes as in (a,b).}
    \label{fig:Neutron}
\end{figure}

\begin{figure}
\includegraphics[width=0.5\textwidth]{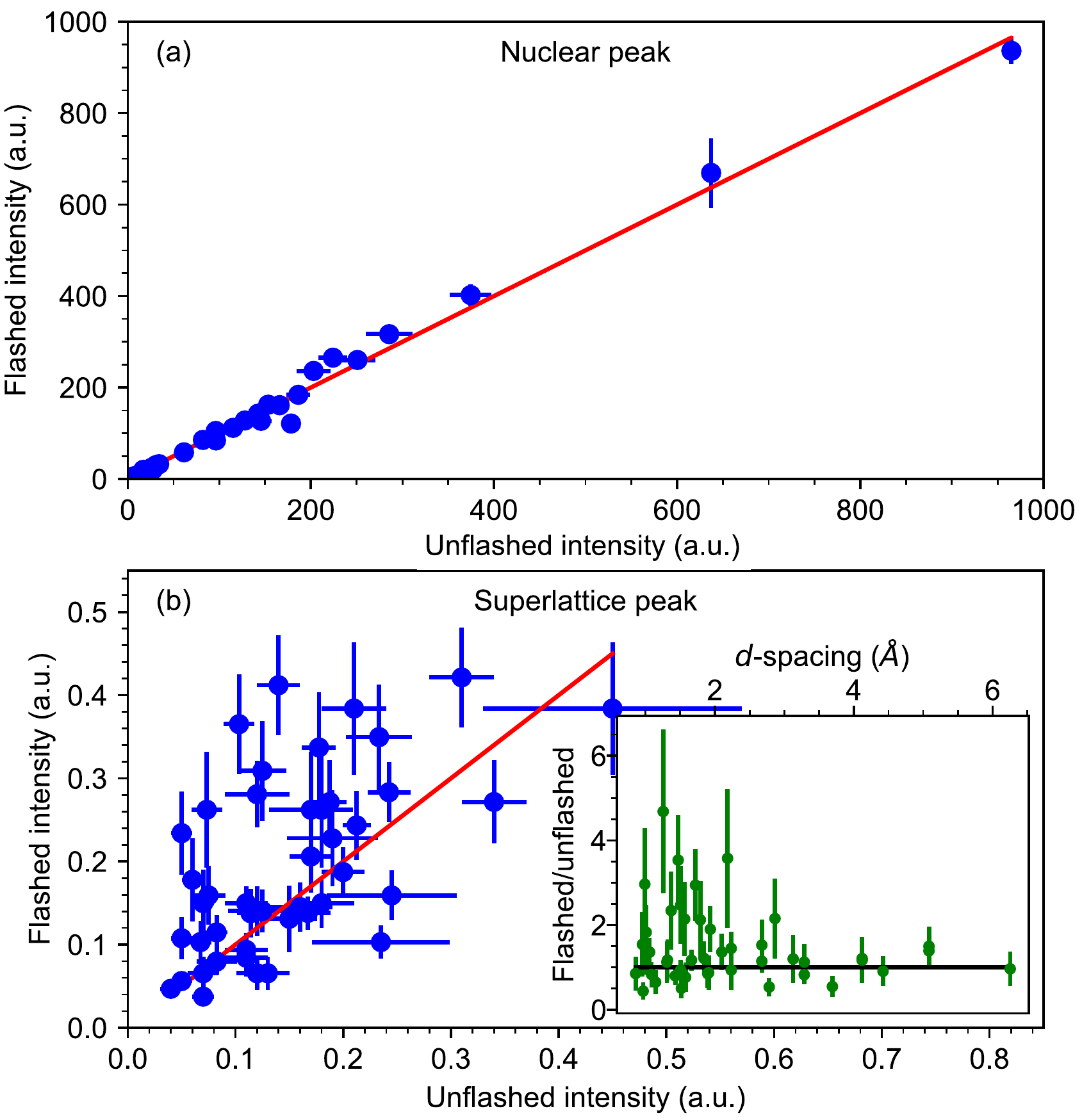}
\centering
    \caption{Comparison of the neutron scattering results for integrated reflections of integer indexed (a) Bragg peaks and (b) superlattice peaks for flashed and unflashed samples. Superlattice intensities were normalized to the integer Bragg peaks to correct for differences in sample volume. Red lines are guides to the eye that correspond to equivalence between the flashed and unflashed samples. Inset in (b) is the ratio of flashed to unflashed superlattice intensities as a function of $d$-spacing, with black line representing unity.}
    \label{fig:intensity}
\end{figure}

\begin{figure}
\includegraphics[width=0.45\textwidth]{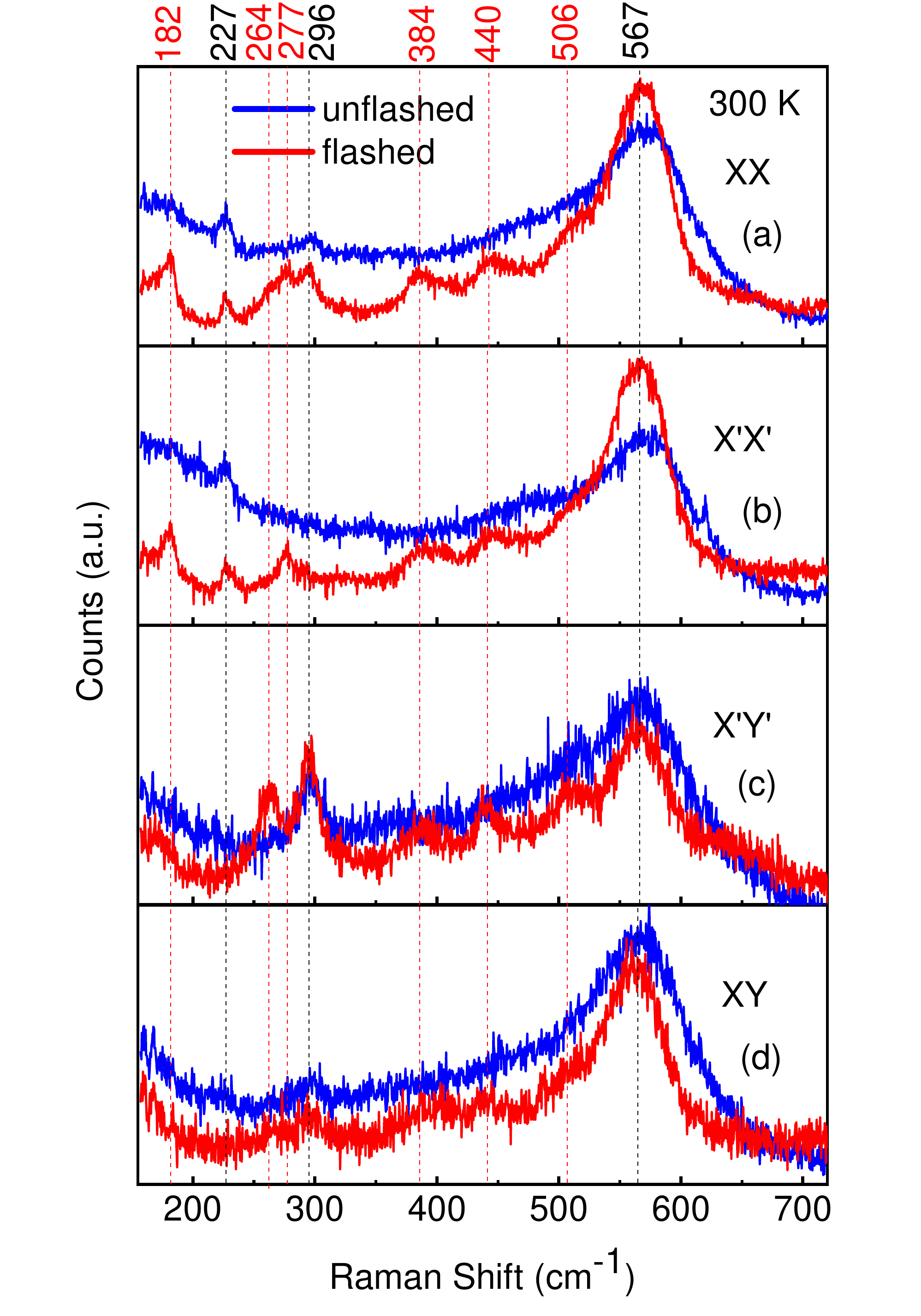}
\centering
    \caption{Raman scattering intensity for the four different polarization geometries (a) XX, (b) X'X', (c) X'Y', and (D) XY for both unflashed and flashed samples. The data were obtained in air at 300 K. The intensities are scaled to match for both samples. The dotted vertical lines and the values on top of the graph indicate the peak positions for the flashed sample. The energies of peaks that are only present in the flashed sample are indicated in red.}
    \label{fig:Polarization_Raman}
\end{figure}


Fig. \ref{fig:Xray} shows that flash had no measurable effect on x-ray Bragg scattering, i.e., both peak positions and intensities were unchanged. Thus the average long-range structure of the two samples is the same within experimental uncertainty. We cannot rule out a small change in oxygen concentration due to the poor sensitivity of x-rays to light atoms such as oxygen.

The neutron scattering cross section of the light oxygen atom is comparable to those of copper and praseodymium, and thus neutron scattering is more accurate and detailed probe of the oxygen positions than x-ray scattering. Fig. \ref{fig:Neutron} shows single-crystal neutron diffraction patterns in the $(hk0)$ and $(hk1)$  scattering planes, indexed based on the $I4/mmm$ 
body-centered unit cell confirmed with x-rays. Similar to the x-ray result, there is no observable effect of flash on the neutron Bragg intensities. This is clearly illustrated in Fig. \ref{fig:intensity}(a), 
which shows a linear correlation between the integrated Bragg peak intensities of the flashed and unflashed samples with negligible deviation. 

In addition to the main Bragg peaks, the neutron diffraction data reveal notably weaker diffraction peaks at fractional indices, \textit{e.g.} $(\sfrac{1}{5},\sfrac{3}{5},0)$, $(\sfrac{4}{5},\sfrac{2}{5},0)$ for $l=0$, and $(\sfrac{1}{2},\sfrac{1}{2},1)$ and $(\sfrac{7}{10},\sfrac{1}{10},1)$ for $l=1$ layer. These peaks can only be indexed in an expanded $\sqrt{10}a\times\sqrt{10}a\times c$ unit cell. The line cuts across the superlattice peaks show resolution-limited widths that are comparable with the average nuclear Bragg reflections [Figs.~\ref{fig:Neutron}(c) and \ref{fig:Neutron}(d)]. The absence of these peaks in the x-ray data implies that they are mainly due to the formation of a super-structure of oxygen atoms that can be understood as oxygen vacancy order in the Pr-O layers. The superlattice peak intensities were integrated and normalized to those of the integer Bragg  peaks in order to correct for differences in sample volume, as seen in Fig.~\ref{fig:intensity}(b). The superlattice peak intensities of the flashed sample tend to be higher, which indicates enhanced oxygen vacancy order. Unfortunately, a detailed refinement of the superlattice was not possible as the supercell is too large.

\begin{figure}
\includegraphics[width=0.5\textwidth]{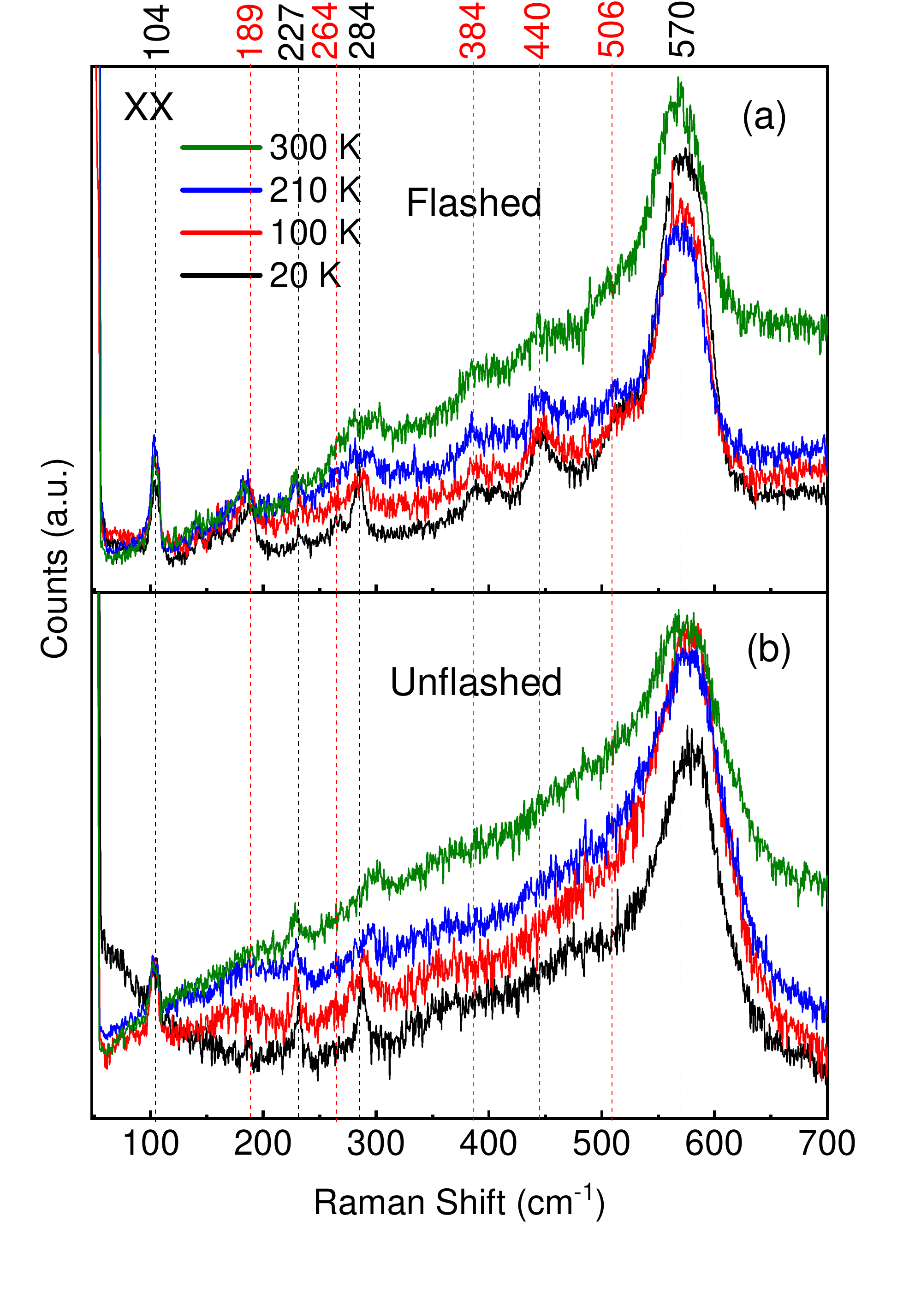}
\centering
    \caption{Temperature dependence of phonons at 20 K, 100 K, 210 K and 300 K for (a) flashed and (b) unflashed samples. XX polarization geometry is used to see all the allowed Raman active modes. The raw data were divided by the Bose factor, and shifted vertically relative to each other for clarity. The dotted vertical lines and the numbers on top of the graph represent energies of the phonon peaks for the flashed sample at 20 K. As in Fig. \ref{fig:Polarization_Raman}, the energies of those peaks that are only present in the flashed sample are indicated in red.}
    \label{fig:Tdependence_Raman}
\end{figure}

In order to further investigate if structural changes occurred due to flash, we used Raman spectroscopy to measure the zone-center phonons (Fig. \ref{fig:Polarization_Raman}, \ref{fig:Tdependence_Raman}). This technique is sensitive to local structure and complementary to x-ray and neutron diffraction. We first present Raman scattering results for the unflashed sample. It is important to note here that published Raman spectra vary significantly among different investigations, and none of them fully matches the spectra that we obtained for both flashed and unflashed samples. Furthermore, many more peaks are observed in the experiments than expected from symmetry considerations. We will present our results here and discuss their implications for understanding the effect of flash further below. Previous investigations \cite{sanjuan1995raman} reported the peak at 228 cm$^{-1}$ as the A$_{1g}$ phonon ($c$-direction vibration of the Pr atoms), and the 303 cm$^{-1}$ peak as the  B$_{1g}$ phonon ($c$-direction vibration of out-of-plane O atoms).  We find the A$_{1g}$ mode at 227 cm$^{-1}$ and the B$_{1g}$ mode at 296 cm$^{-1}$. A broad enigmatic peak near 550 cm$^{-1}$ is seen in all polarizations consistent with previous observations\cite{sugai1990magnon,sanjuan1995raman,sanjurjo1994raman}. In our experiments, this peak is centered at 570 cm$^{-1}$ and has a shoulder around 465 cm$^{-1}$.  Similarly to previous reports, we do not observe any A$_{2g}$ or B$_{2g}$ phonon or electronic signal in this sample. 

Extra peaks appear in all polarization geometries in the flashed sample (Fig. \ref{fig:Polarization_Raman}). In X'X' geometry, there are two new peaks of A$_{1g}$ symmetry at 182 cm$^{-1}$ and 277 cm$^{-1}$. For X'Y', the single B$_{1g}$ phonon mode in the unflashed sample, splits into two peaks at 264 and 296 cm$^{-1}$ in the flashed sample. This split peak belongs to the B$_{1g}$ symmetry as it is present in both XX and X'Y' polarizations. At higher energies, the broad peak at 570 cm$^{-1}$ in the unflashed sample softens to 567 cm$^{-1}$ after flash. Moreover, this peak is more intense and narrower in the flashed sample. Finally, we  observe three new peaks at 384, 440, 506 cm$^{-1}$. 

Figure \ref{fig:Tdependence_Raman} illustrates that temperature has only a small effect on the phonons above 350 cm$^{-1}$ and on the modes at 104 cm$^{-1}$, 227 cm$^{-1}$, and 264 cm$^{-1}$. The lower energy A$_{1g}$ phonon (227 cm$^{-1}$)  softens slightly and the B$_{1g}$ phonon (303 cm$^{-1}$) hardens with temperature in the unflashed sample. The 284 cm$^{-1}$ peak hardens to 294 cm$^{-1}$. 

We now focus on the peaks between 250 and 300 cm$^{-1}$ in  the flashed sample. At 300K (Fig. \ref{fig:Polarization_Raman}) there are two peaks of B$_{1g}$ symmetry (264 and 296 cm$^{-1}$) and one is A$_{1g}$ (277 cm$^{-1}$). The latter is absent at 20K where only the two B$_{1g}$ peaks appear. Thus we assign the A$_{1g}$ peak to a crystal-field excitation and the B$_{1g}$ peaks to phonons. These three peaks together form a broader peak at 210 K in the flashed sample. Note that the 264 cm$^{-1}$  peak does not change with decreasing temperature, whereas the 296 cm$^{-1}$ peak softens to 284 cm$^{-1}$ at 20 K. We did not observe any significant  differences between the flashed and unflashed samples in the ZZ polarization geometry, and the spectra agree with earlier results obtained on undoped sample \cite{sanjuan1995raman}.

\begin{figure}
\includegraphics[width=0.45\textwidth]{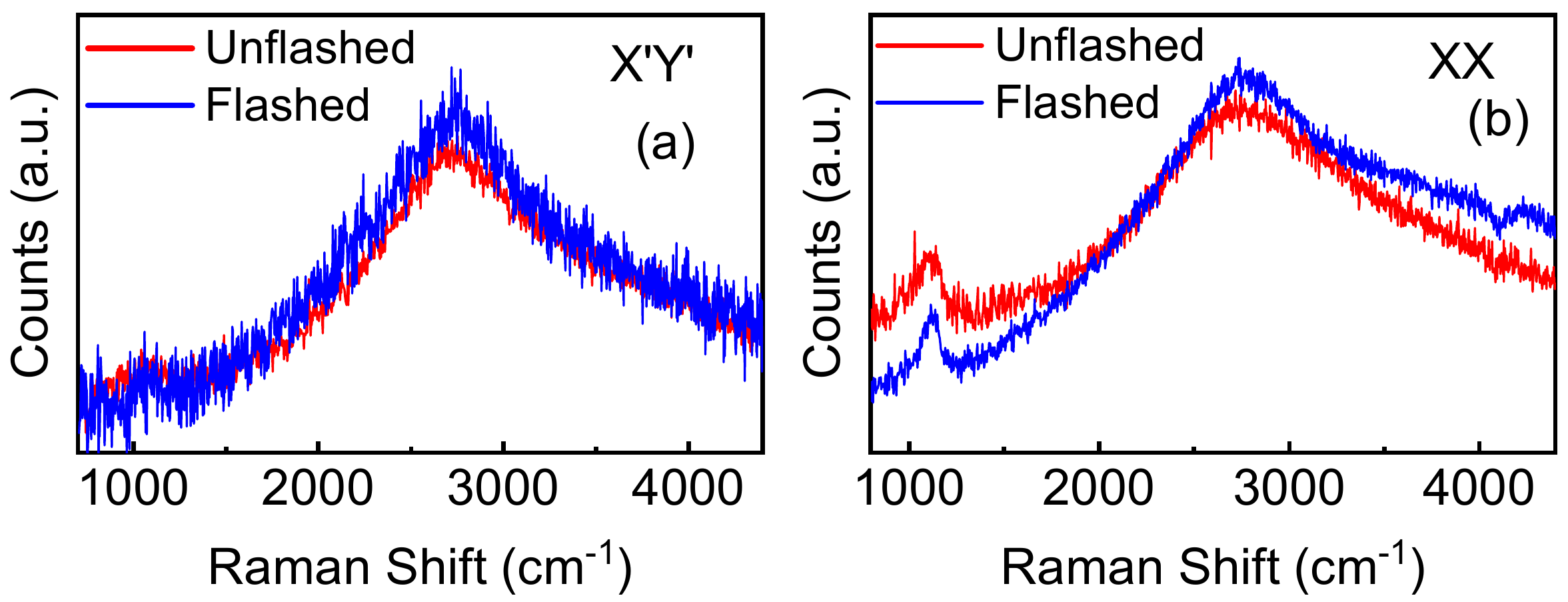}
\centering
    \caption{Comparison of the two-magnon peaks of the unflashed and flashed samples. The data were shifted vertically relative to each other and different scaling factors were applied for ease of comparison in X'Y' (a) and XX (b) polarization geometries.}
    \label{fig:2M}
\end{figure}

At even higher energies there is a peak in XX geometry at 1128 cm$^{-1}$ that was previously identified as two-phonon scattering. It is narrower in the flashed sample (Fig. \ref{fig:2M}).  

The effect of flash in the two-dimensional antiferromagnetic interactions in the CuO$_2$ plane can be deduced by examining the broad two-magnon Raman peak at 2700 cm$^{-1}$ (Fig. \ref{fig:2M}), which appears in X'Y' and XX polarization geometries because of its B$_{1g}$ symmetry \cite{sugai1990magnon}. Two-magnon Raman scattering at the peak maximum corresponds to two neighbouring spins coupled by the exchange interaction \emph{J} that are flipped through virtual charge-transfer excitations.  We find that the linewidth and position of the peak remain almost unchanged by flash (Fig. \ref{fig:2M}).

\section{Discussion}

In principle,  the D$_{4h}^{17}$ crystal symmetry allows Raman-active modes of A$_{1g}$ + B$_{1g}$ symmetry in XX, A$_{1g}$ + B$_{2g}$ in X'X', A$_{2g}$ + B$_{1g}$ in X'Y', and A$_{2g}$ + B$_{2g}$ in XY polarizations. However, the nearly stoichiometric structure of PCO allows only four modes: 2E$_g$, 1A$_{1g}$, and 1B$_{1g}$, all of which involve vibrations of Pr or interplane oxygen (on O$_2$-sites). Only two modes, A$_{1g}$ and B$_{1g}$ can be observed in the scattering geometries of our experiment \cite{HEYEN19901299}. However more than two modes are present in our data as well as in all other published Raman data for PCO presumably because of disorder and crystal field excitations \cite{sanjurjo1994raman,sanjuan1995raman}. 

Since Raman scattering is a bulk probe, and an impurity phase \cite{PhysRevB.70.094507} contributes to the phonon intensity, our data would seem to imply a large impurity volume fraction if the extra phonon peaks originated from and impurity phase. In this case impurity-phase Bragg peaks would be expected in the diffraction data. However, we did not observe any Bragg peaks that do not originate from PCO. It is therefore unlikely that the impurity phase contributes significantly to the Raman phonon intensity.

Nominally stoichiometric perovskites can exhibit inherent structural inhomogeneity as well as disorder due to excess oxygen or oxygen vacancies [e.g. \cite{Pelc2022}, and references therein]. It is known that oxygen-vacancy order can lead to novel electronic properties [e.g., \cite{PhysRevMaterials.2.111404}]. Our work uncovers such oxygen-vacancy disorder in both as-grown and modified Pr$_2$CuO$_4$, along with a significant effect of flash on the degree of this order, thereby opening a new arena for the manipulation and study of the interrelation between structural and electronic properties in a prominent class of quantum materials.

We believe that the unexpected phonon peaks observed by Raman scattering originate from the presence of oxygen vacancies. Some of these vacancies are ordered into a superlattice, whereas others are not. Raman-forbidden phonons can become Raman active if the crystal symmetry is locally disturbed by these ordered or unordered vacancies. The energies of the new peaks in the flashed sample do not match those of the IR modes \cite{homes2002infrared}, so the extra peaks likely stem from a folding of the phonon branches due to the presence of the superlattice. Line-broadening is primarily present as a result of disorder due to unordered vacancies. Differences in concentration and ordering of these vacancies can explain discrepancies between the Raman data by different research groups.

Our diffraction data show that the superlattice peaks become more intense after the flash/quench treatment, i.e., that vacancy-order increases. Based on the proposed scenario, increased ordering of oxygen vacancies should sharpen the phonon peaks, as observed in experiment, as the degree of disorder is reduced.

The atomic vibrations of the CuO$_2$ layers are all Raman-forbidden, so phonon Raman scattering probes Pr-O layers only. The dramatic change of the phonon spectrum as a result of flash therefore reflects increased oxygen-vacancy order that, according to refinement of the diffraction data, exclusively involves the O sites of the Pr-O layers. 

Two-magnon scattering, on the other hand originates from moments on the Cu sites, i.e., it directly probes the CuO$_2$ layers. The two-magnon scattering is observed to be the same in the flashed and unflashed samples, consistent with the picture that only the Pr-O layers are influenced by flash.

We also found that some bulk properties are strongly affected by flash, whereas others are not. Properties associated with the CuO$_2$ planes, such as in-plane magnetic susceptibility, ab-electrical resistivity, the magnetic exchange constant \emph{J} (proportional to the energy of the two-magnon peak), and doping (which determines the two-magnon peak intensity), are not affected within experimental uncertainty. On the other hand, properties dependent on the Pr-O layers, such as the $c$-axis magnetic susceptibility and the low-temperature $c$-axis electrical resistivity are significantly altered leading to an increased three-dimensional nature of the charge transport and likely also magnetism.


We conclude that passing electrical currents large enough to induce a nonlinear response followed by a LN$_2$ quench, is a promising new way to modify and control properties of correlated perovskite oxides potentially leading to new functionality. 


\section{Acknowledgement}
S.R. and D.R. who performed Raman scattering, bulk measurements, and  flash experiments were supported by U.S. Department of Energy, Office of Basic Energy Sciences, Office of Science, under Contract No. DE-SC0006939. The work at the University of Minnesota was funded by the U.S. Department of Energy through the University of Minnesota Center for Quantum Materials, under Grant No. DE-SC0016371. The work of S.I.J and R.R. who performed and designed flash experiments and helped with interpretation of the data were supported by the Office of Naval Research under grant N00014-18-1-2270. We thank Dr. Antti Makinen for taking an interest in this project. A portion of this research used resources at the Spallation Neutron Source, a DOE Office of Science User Facility operated by the Oak Ridge National Laboratory.
\bibliography{PCO}

\end{document}